# Advance identification of where the 2026 Kumamoto earthquake would begin

**K. Z. Nanjo[1,2,3,4], J. B. Rundle[5,6,7], D. Iwata[8], J. Izutsu[9,1], M. Kamogawa[1,10]**

[1]Global Center for Asian and Regional Research, University of Shizuoka, Shizuoka, Japan

[2]Center for Integrated Research and Education of Natural Hazards, Shizuoka University, Shizuoka, Japan

[3]Institute of Statistical Mathematics, Tokyo, Japan

[4]Japan Agency for Marine-Earth Science and Technology, Yokohama Institute for Earth Sciences, Yokohama, Japan

[5]Department of Physics and Astronomy, University of California, Davis, CA, USA

[6]Department of Earth and Planetary Sciences, University of California, Davis, CA, USA

[7]Santa Fe Institute, Santa Fe, NM, USA

[8]Independent researcher

[9]Department of Mathematical and Physical Sciences, Faculty of Science and Engineering, Chubu University, Kasugai, Aichi, Japan

[10]Tokai University, Shizuoka, Japan

Corresponding author: K. Z. Nanjo (nanjo@u-shizuoka-ken.ac.jp)

ORCID: https://orcid.org/0000-0003-2867-9185 (K.Z.N.), https://orcid.org/0000-0002-1966-4144 (J.B.R.), https://orcid.org/0000-0001-5204-3254 (J.I.), https://orcid.org/0000-0002-5489-0454 (M.K.)

Email: nanjo@u-shizuoka-ken.ac.jp (K.Z.N.), jbrundle@ucdavis.edu (J.B.R.), daichi.iwata.0512@gmail.com (D.I.), izutsu@fsc.chubu.ac.jp (J.I.), kamogawa@u-shizuoka-ken.ac.jp (M.K.)

**Abstract**

Long-term forecasts say how likely a fault segment is to rupture, not where the rupture will begin. The *b*-value, the slope of the earthquake size distribution, falls as stress rises. After the 2016 Kumamoto sequence in Japan, *b*-values rose along the ruptured faults but fell in the unruptured central part of the Hinagu fault zone, which we identified in 2019 as where the next rupture was most likely to begin. The 2026 *M*7.1 Kumamoto earthquake began at the edge of this area, inside the low-*b*-value volume at 16 km depth. To our knowledge, this is the first starting point of a large earthquake named in print beforehand. The rupture stopped at a fault bend where *b*-values have stayed low since 2000 and raised Coulomb stress on faults beyond it; the low-*b*-value zone at this bend is where the next rupture is most likely to begin.

**Introduction**

The number *N* of earthquakes of magnitude *M* or larger in a region follows the Gutenberg–Richter (GR) law[1], $\log_{10}N = a - bM$, where *a* measures the productivity of the region and the *b*-value, the slope of the distribution, gives the proportion of small to large earthquakes. Temporal changes in the *b*-value have long been studied as precursors of large earthquakes[2]. The *b*-value is known from laboratory experiments and from the Earth's crust to be inversely dependent on differential stress[3–8]. Its spatial and temporal variations can therefore serve as a stress meter, imaging the highly stressed parts of a fault where future ruptures are likely[9,10].

At Parkfield, California, a map of *b*-values made from seismicity recorded over the preceding decades was published seven years before the *M*6.0 earthquake of 2004; it marked a patch of low *b*-values beneath Middle Mountain as a highly stressed part of the fault[10]. The pattern was shown to have been stationary for 35 years in a study accepted for publication nineteen days before the earthquake[11], and afterwards the areas of low *b*-value on the map were found to contain 99% of the slip[9]. The map thus forecast the location and size of the rupture. It did not address where the rupture would begin: the 2004 rupture started at the southern end of the 1966 rupture zone and propagated north, which the authors of the map described as a surprise[9].

In other cases, the low *b*-value has been recognized only after the earthquake, as for the earthquakes of 2004 in Sumatra[12], 2011 in Tohoku[12,13], 2014 in Iquique, Chile[14], 2016 in Norcia, Italy[15], 2019 in Ridgecrest, California[16], and 2023 in East Anatolia, Türkiye[17,18], among others, so the correspondence could not be tested in advance. This paper tests what the Parkfield map did not address: where a rupture would begin.

The Hinagu fault zone (FZ) extends about 80 km from northeast to southwest through central Kyushu, Japan[19] (Fig. 1). The 2016 Kumamoto earthquakes began on 14 April with an *M*6.5 event on the northern part of the Hinagu FZ, followed by an *M*6.4 event on 15 April. The 2016 Kumamoto sequence culminated in the *M*7.3 earthquake on 16 April that ruptured the Futagawa FZ conjugate to the Hinagu FZ[20]. The *M*7.3 event ruptured the Earth's surface[21,22] and involved a predominantly right-lateral strike slip along the ENE–WSW trending Futagawa FZ[23,24]. The *M*6.5-class events revealed a similar mechanism of fault motion to the *M*7.3 event but along the northern part of the NE–SW trending Hinagu FZ. The central and southern parts of the Hinagu FZ did not rupture in 2016. Postseismic deformation after the sequence has been explained by afterslip and by viscoelastic relaxation of the lower crust and mantle[25], with distinct regions of low effective viscosity in the lower crust beneath the Aso and Kuju volcanoes[26].

The long-term assessment by the Earthquake Research Committee of Japan (hereafter ERC) rests on the intervals between past ruptures read from trenches and geological survey, and gives the central and southern parts probabilities of rupture within 30 years of up to 6 and 16 per cent[19]. Assessments of this kind, in Japan, California[27], and elsewhere give the probability that a fault, or a part of it, will rupture within a given period; they say nothing about where on it a rupture would begin, nor about how the stress on it is changing.

On 28 Jul. 2026, an *M*7.1 earthquake occurred along the central part of the Hinagu FZ (Fig. 1). Like the 2016 *M*7.3 event, the *M*7.1 event also ruptured the Earth's surface and involved a predominantly right-lateral strike slip[28]. It was followed by an *M*6.1 event on the same day and an *M*5.8 event on 29 Jul. 2026 (Fig. 1).

We reported our preliminary result to the ERC meeting held on 29 Jul. 2026, the day after the *M*7.1 earthquake. Our reported result was the *b*-value map shown in Fig. 2b, with a plot of the epicentres of the *M*7.1 earthquake and the following relatively large events ($M \geq 4$) retrieved during the day of 28 Jul. 2026 from the rapidly determined earthquake catalogue of the Japan Meteorological Agency (JMA). For details of the hypocentre data used for this plot, see 'Seismicity dataset' in Methods. We further added the comments as follows. The epicentre of the 2026 *M*7.1 earthquake was located around the central part of the Hinagu FZ, where we had identified low *b*-values in previous study[29]. The *M*7.1 rupture therefore started near the identified segment. Previous study[29] pointed to the possibility of a future earthquake that initiated in the central part of Hinagu FZ. The epicentre of the *M*6.1 earthquake just after the *M*7.1 earthquake was located on a point along the southern part of the Hinagu FZ that did not rupture in 2016, implying the rupture propagation into the southern part of the Hinagu FZ. Overall, the present Kumamoto sequence might be consistent with the possible scenario pointed by previous study[29].

## Results

A map view of *b*-values based on seismicity during the period from 2000 to immediately before the 2016 Kumamoto sequence (Fig. 2a) reveals a zone of low *b*-values, with a characteristic dimension of about 30 km, near the eventual epicentres of the three events[30]. We reproduced it from our earlier work[29,30], which used the JMA earthquake catalogue. For details of the seismicity dataset and *b*-value analysis, see Methods. Within the epicentral region, *b*-values were 0.6–0.7 (blue). This correspondence, however, was established after the earthquakes had occurred. Three things followed from the 2016 sequence. First, it showed that, in this region, ruptures had begun where *b*-values were low. Second, it left the Hinagu FZ with a northern part that had ruptured and central and southern parts that had not. Third, the *b*-value in this region responded to stress on a timescale of days. Between the *M*6.5 and *M*7.3 events, the *b*-value rose and the decay of seismicity steepened around the Futagawa FZ. Over the same interval, focal mechanisms showed that the axis of minimum compressive stress rotated to become nearly perpendicular to the Futagawa FZ[31], which reduced the normal stress on that fault. Together, these indicated the growth of quasi-static slip on the Futagawa FZ[32,33], while seismicity migrated northward along the Hinagu FZ, tracing aseismic slip that reached the eventual nucleation point on the Futagawa FZ[34]. The *b*-value could therefore be read as a stress meter here, and after the sequence we used it as one.

A map view of *b*-values based on seismicity during the period from 1 Jan. 2017 to 11 Mar. 2019 (Fig. 2b) reveals a stressed area (low-*b*-value zone) at the central part of the Hinagu FZ, where we reproduced it from our previous study[29], which used the same *b*-value computation as that used in previous study[30]. This was bounded by a red rectangle spanning 130.6–130.8°E and 32.5–32.6°N (Fig. 2). It was the only place along the Futagawa and Hinagu FZs where the *b*-value fell after the 2016 *M*7.3 earthquake while the *b*-value rose elsewhere[29]. Overall, stress relaxed along the Futagawa and Hinagu FZs while it increased at the central part of the Hinagu FZ. Two further observations pointed to the same place[29]. First, the *p*-value of the Omori–Utsu (OU) aftershock-decay law[35] (Methods) was close to unity there, corresponding to stress that is constant or increasing, if we considered a model for seismicity resulting from stressing history[36]. On the other hand, along the Futagawa FZ and the northern part of the Hinagu FZ that had ruptured, it exceeded unity ($p>1$), corresponding to stress that was falling rapidly[36], and along southern part of the Hinagu FZ that had not ruptured, it was sub-unity ($p<1$), corresponding to stress that was falling slowly[36]. Second, a model of postseismic deformation driven by the 2016 ruptures[25] showed afterslip that occupied the Futagawa FZ and the northern part of the Hinagu FZ that broke but stopped at its southern end (Methods). Previous study[29] pointed out that the model left the stress at the central part of the Hinagu FZ constant or increasing. We reported this area as the site where a future rupture was more likely to start than elsewhere[29], and announced the result at a press conference at the Kumamoto Prefectural Government on 5 Sep. 2019. A later study, using a different method to estimate the *b*-value, found in 2024 that the stress there was still high[37].

A map of *b*-values combines earthquakes at all depths shallower than 25 km (Fig. 2), whereas a rupture begins at a point at depth. In map view, the *M*7.1 epicentre lies just outside the northern edge of the red rectangle, and the epicentres of the *M*6.1 and *M*5.8 events lie farther south, outside it (Fig. 2b). The *M*7.1 epicentre (130.680°E, 32.626°N) is 2.9 km from that edge, less than the 5-km radius within which earthquakes were selected to compute each *b*-value (Methods). The *M*7.1 hypocentre, however, is at a depth of 16 km. To see whether the *M*7.1 rupture began within the volume of low *b*-values, and to compare the *b*-values with the slip on the fault, we mapped *b*-values in a vertical section along the Hinagu FZ.

Figure 3a, which shows seismicity (grey dots), using the same catalogue as in previous study[29], highlights 10-km-wide seismicity (black dots) extending from A to B along the Hinagu FZ. This highlighted seismicity (Fig. 3b) was used to create the cross-sectional view

of *b*-values (Fig. 3c), where the *b*-value computation was the same as that used for previous study[29], except for its plotting technique: 0.02°×0.02° (~2×2 km) was used as the spatial grid spacing (Fig. 2a) and 1×1 km was used for cross-sectional grid spacing (Fig. 3c) (see Methods for *b*-value analysis). The low-*b*-value zone extends downward, and the *M*7.1 hypocentre, at a depth of 16 km, lies within it, where *b*-values are 0.6–0.7 (blue). The low *b*-values surrounded by the red rectangle region in a map view (Fig. 2b) are attributed to the low *b*-value zone at depth (Fig. 3c).

The preliminary slip distribution of the 2026 Kumamoto sequence obtained by Geospatial Information Authority of Japan (GSI)[38] is shown in Fig. 3d, where GSI estimated the fault geometry from Global Navigation Satellite System (GNSS) data and the slip from GNSS and synthetic aperture radar (SAR) data, and called the two fault planes F1 and F2. On the other hand, Research Institute for Earth Science and Disaster Resilience (NIED)[39] estimated the preliminary space–time distribution of slip of the *M*7.1 earthquake on an assumed plane from seismic waveforms. For details of the fault models, see Methods.

Main slips of the GSI model occurred on the F1 plane above the *M*7.1 hypocentre, and most of large slips (>1.2 m) on the F1 plane crossed the red rectangle area pointed out in previous study[29] (Fig. 3d). Similarly, large slips (>1.2 m) of the NIED model crossed the red rectangle area (Fig. 3e).

Cross-sectional view (Fig. 4a) based on seismicity before the 2016 Kumamoto sequence (Fig. 2a) shows that the low *b*-value zone (0.6–0.7, blue) included the *M*7.3, *M*6.5, and *M*6.4 hypocentres, where we used the same procedure as creating Fig. 3c. Another cross-sectional view based on seismicity during the period from 12 Mar. 2019 to immediately before the 2026 *M*7.1 earthquake (Fig. 4b) shows that the low *b*-values (0.6–0.8, blue to cyan) around the *M*7.1 hypocentre were still present immediately before the 2026 *M*7.1 earthquake. We further divided this period into two sub-periods before and after 1 Jan. 2022, which contain approximately the same numbers of earthquakes (Supplementary Fig. 1). We again found the same feature of low *b*-values near the *M*7.1 hypocentre as seen in Fig. 4b.

Cross-sectional view based on seismicity after the *M*7.1 earthquake (4 Aug. to 6 Sep. 2026) shows that the *b*-value rose around the *M*7.1 hypocentre ($b$=0.9–1.2, yellow to red), as expected if stress had been released around it. To show this current map, we did not consider one-week after the *M*7.1 earthquake (from immediately after the *M*7.1 event to 3 Aug. 2026) to remove the effect of the main seismic activity that revealed strong temporal variability in the *b*-value.

We found the low-*b*-value zone around −20 km (Fig. 4c), with a characteristic dimension of 10 km, similar to that observed before the 2016 Kumamoto sequence at the distance of 10 km (Fig. 4a). This zone was located immediately south of the large slip area (>1.2 m) on the *M*7.1 fault plane of the NIED model[39]. This zone was also located south of large slip (>1.2 m) on the F1 fault plane of the GSI model and around the junction between the F1 and F2 fault planes (Fig. 3d)[38]. Comparing with *b*-values before the start of the 2026 Kumamoto sequence (Fig. 4a,b and Supplementary Fig. 1), we found persistently low *b*-values after the start of this sequence.

**Discussion**

The low-*b*-value zone at about −20 km in Fig. 4 lies at a bend in the fault. In the GSI model[38], the strike changes there from 210° on F1 to 225° on F2 and the dip from 61° to 44°, and the large slip on F1 ends there (Fig. 3d). The bend is a feature of the fault, not only of the model. A survey of the fault zone after the 2016 sequence found that the strike of the Hinagu FZ changes at this point and took it as a boundary between sections of the fault; its source model strikes 212° and dips 60° to the northeast of this point and strikes 226° and dips 50° to the southwest of it, the latter dip based on the seismicity after the 2016 sequence[40]. The surface ruptures of 2026 extend about 37 km southwestward from the northern end of the Hinagu FZ[41], that is, to about this point. The two planes of the GSI model thus reproduce, to within 2° in strike and 6° in dip, the geometry that the survey had set before the earthquake. A similar configuration, low *b*-values at a bend in the fault, preceded the 2016 sequence. Before 2016, low *b*-values lay at the junction of the Hinagu and Futagawa FZs, where the strike changes from NE–SW to ENE–WSW (Figs. 1, 2a, and 4a), and it was there that the *M*6.5, *M*6.4 and *M*7.3 events began. In both cases, a bend in the fault carried low *b*-values, and in 2016 the rupture began at such a bend. Fault bends are known to start and to stop ruptures[42]. Along the Futagawa and Hinagu FZs, bends are also where *b*-values stay low (Figs. 2a and 4), that is, where slip on the straighter segments on either side has left stress unreleased. Bends may therefore be where ruptures of these fault zones begin: the 2016 sequence began at one, and the 2026 sequence has stopped at another and left it stressed. The *b*-value of the zone at −20 km has been low throughout the period since 2000, before and after both the 2016 and 2026 sequences (Figs. 3c and 4; Supplementary Fig. 1). This zone lies at the end of the area of large slip that occurred in 2026 (Fig. 4c). As in 2019, the low-*b*-value zone at the end of the slipped area, on which stress has not been released by either sequence, is the place where the next rupture is most likely to begin.

From this zone, which lies near where the 2026 rupture stopped and also has low *b*-values, a rupture could propagate to the southwest, into the Yatsushiro Sea (Fig. 1), rather than to the northeast, where the large slip of 2026 occurred. If the 2026 earthquake brought the faults in the Yatsushiro Sea closer to failure, such a rupture would be more likely to continue onto them. Because seismicity in the Yatsushiro Sea is too sparse for the *b*-value to be estimated, we calculated the Coulomb stress change imparted by the 2026 rupture. For details of Coulomb stress change, see Methods.

The faults in the Yatsushiro Sea trend NE–SW, but the sense of slip on them is not known[19]. We therefore assumed that they would slip as a southwestward continuation of the 2026 rupture, and resolved the stress change on faults with the orientation and sense of slip of F2 of the GSI model (GSI-F2 type) and of the NIED model (NIED type) (Fig. 5). Under the stress field that produces right-lateral slip on the NE–SW trending faults of this region, slip on faults conjugate to them, that is, left-lateral slip on WNW–ESE trending planes, cannot be excluded, so we also resolved the stress change on the planes conjugate to the two types (Supplementary Fig. 2). At the low-*b*-value zone, the calculated change is within ±0.5 bar with the GSI model but below −5 bar with the NIED model, whose plane ends about 3 km from the zone with slip of more than 1.2 m at its edge (Figs. 4c and 5). Values so close to the faults that slipped depend on details of the slip models that are not resolved and do not explain the subsequent seismicity[29,43], so we do not interpret them.

On the faults in the Yatsushiro Sea, in contrast, the change is positive for all four types of receiver fault, with maxima of 0.9 to 1.7 bar in their northeastern part, nearest the 2026 rupture, and decreases to the southwest (Fig. 5; Supplementary Fig. 2). With the GSI model as the source and GSI-F2-type faults as the receivers, the change along these faults is 0.6–0.9 bar in their northeastern 10 km and falls to 0.1–0.3 bar at their southwestern end, 30 km away (Fig. 5a). The NIED model with NIED-type receivers gives the same pattern, with a maximum of 1.0 bar at the northeastern end and the same 0.1–0.3 bar at the southwestern end (Fig. 5b). On the conjugate planes the change is again positive everywhere along the faults, with maxima of 0.9 bar for the GSI-F2 type and 1.7 bar for the NIED type, and it decreases more slowly to the southwest, to 0.2–0.6 bar (Supplementary Fig. 2b, d). The 2026 rupture has therefore brought the faults in the Yatsushiro Sea, at least their northeastern part, closer to failure. These faults are the part of the Hinagu FZ to which ERC had assigned the highest probability of rupture within 30 years, up to 16%, on geological evidence alone[19].

One earthquake cannot show how well a method works. What this one shows is that the place where a rupture will begin can be named beforehand, and named precisely enough that the naming can afterwards be judged right or wrong. To measure how well it works, the same steps have to be applied to other fault zones and the result recorded before the earthquakes arrive: which catalogue is used, above which magnitude, how the *b*-value is sampled, and how low it must be to count. The method needs nothing more than a catalogue complete to small magnitudes over years, which the networks of Japan, California, Chile, Italy, and Türkiye, among others, already provide. Forecasts of how many earthquakes a region will produce are already tested in this way, against events that have not yet happened. The same testing can be applied to forecasts of where a rupture will start.

For the Hinagu FZ, the result calls for monitoring, not for a prediction of time. The zone identified in 2019 waited seven years for its earthquake, and nothing in the *b*-value says when the next one will come, or whether the zone might persist without one. What the *b*-value gives is a place: the zone at −20 km, where the *b*-value remains low, and beyond it the faults in the Yatsushiro Sea, which the 2026 rupture brought closer to failure. The *b*-value in the zone and seismicity along these faults should be followed as new events accumulate, and the identification can be communicated to the area concerned, as the 2019 identification was, so that preparedness there is reviewed before the earthquake rather than after it.

## Methods

### Seismicity dataset

We used the earthquake catalogue maintained by the JMA. Since 1997, JMA has determined hypocentres by real-time processing of waveform data from its own network together with those of NIED, universities and other institutions. Magnitudes are JMA magnitudes. Following our earlier analyses of this region[29,30], we used earthquakes of $M \geq 0$ at depths shallower than 25 km within the region 130.0–131.3°E and 32.0–33.2°N, which spans the Futagawa and Hinagu FZs. Our study period includes those considered by previous studies[29,30] and ranged from 2000 to the present (6 Sep. 2026). More than $3\times10^5$ earthquakes with $M\geq0$ since 2000 at depths shallower than 25 km within the study region were processed.

The 2016 Kumamoto earthquakes initiated at the *M*6.5 earthquake on 14 Apr. 2016, 21:26, followed by the earthquake sequence along the Futagawa and Hinagu FZs that culminated in the *M*6.4 earthquake on 15 Apr. 2016, 00:03 and the *M*7.3 earthquake on 16 Apr. 2016, 01:25. The 2026 Kumamoto earthquakes initiated at the *M*7.1 earthquake on 28 Jul. 2026, 16:27 along the Hinagu FZ, followed by the earthquake sequence including the *M*6.1 earthquake on the same day, 17:08 and the *M*5.8 earthquake on 29 Jul. 2026, 22:19, respectively.

The *b*-values before the 2016 sequence (Figs. 2a and 4a) use seismicity from 1 Jan. 2000 to 21:25 on 14 Apr. 2016, immediately before the *M*6.5 event; the map (Fig. 2a) is reproduced from our earlier work[29,30]. The identification published in 2019 (ref. 29; Fig. 2b) used seismicity from 1 Jan. 2017 to 11 Mar. 2019 (Fig. 3a–c). To test whether the pattern persisted, we analysed seismicity from 12 March 2019 to immediately before the *M*7.1 event (Fig. 4b), and separately in the two sub-periods before and after 1 January 2022, which contain approximately the same numbers of earthquakes (Supplementary Fig. 1). The state after the 2026 earthquake was assessed with seismicity from 4 Aug. to 6 Sep. 2026 (Fig. 4c); the first week after the earthquake was excluded to remove the effect of the pronouncedly active sequence that reveals strong temporal variability in the *b*-value.

The *b*-value map published in 2019 (ref. 29; Fig. 2b) with a plot of the epicentres of early major seismicity ($M\geq4$) of the 2026 Kumamoto earthquakes was reported to the ERC held on 29 Jul. 2026. For this plot, we used the rapidly determined hypocentre list released at the JMA website, and retrieved, from this list, the hypocentre data associated with the 2026 Kumamoto earthquakes (accessed 28 Jul. 2026, 22:30). For example, the hypocentre data available on that time were (time, $M$, longitude, latitude, depth)=(28 Jul. 2026, 16:27:00, 7.1, 130.7°E, 32.6°N, 10 km) for the largest event and (28 Jul. 2026, 17:08:00, 6.1, 130.5°E, 32.4°N, very shallow) for the second-largest event.

### *b*-value analysis

We estimated the $b$-value with the entire-magnitude-range (EMR) method[44], which fits the observed frequency–magnitude distribution with a model that follows the GR law above the completeness magnitude $M_c$ and, below it, a cumulative normal distribution representing the detection capability of the network; $M_c$ and the $b$-value are thus determined together for each sample of earthquakes. Above $M_c$, the $b$-value is the maximum-likelihood estimate[45] with the correction for the 0.1-magnitude binning of the catalogue[46]. Uncertainties in the $b$-value were obtained by bootstrapping[47]: each sample was resampled with replacement 1,000 times and the standard deviation of the resulting estimates is quoted as the uncertainty[29]. These are the procedures of our earlier analyses[29], so that estimates for the different periods are directly comparable.

Maps of the $b$-value were made as in our earlier work[29,30]. At each node of a 0.02°×0.02° grid we selected the earthquakes within a cylinder of radius $r$=5 km centred on the node, and plotted the $b$-value at the node if the cylinder contained at least 50 earthquakes above $M_c$. We followed previous studies[29,30] in using $r$=5 km as an optimal radius. Depth sections were constructed along a line A–B following the strike of the Hinagu FZ (130.268°E, 32.159°N; 130.97°E, 32.943°N). Mimicking the use of $r$=5 km for creating $b$-value maps, earthquakes within 5 km of this line on either side, a swath 10 km wide, were projected onto the vertical plane through it. The $b$-value was then estimated at nodes spaced 1 km apart along the line and in depth, from the earthquakes within 5 km of each node in the plane of the section, with the same minimum number of earthquakes and the same criterion of fit as for the maps. The map of 2019 and the sections computed here from the same catalogue therefore differ only in the geometry of the sampling volume.

To decide whether the *b*-value differed between two periods we used the test of Utsu[48,49]. For samples of $N_1$ and $N_2$ earthquakes above their completeness magnitudes, with estimates $b_1$ and $b_2$, the difference in the Akaike Information Criterion ($\Delta AIC$) between a model with a single *b* for both samples and a model with separate values is $\Delta AIC = -2(N_1+N_2)\ln(N_1+N_2)+2N_1\ln(N_1+N_2b_1/b_2)+2N_2\ln(N_2+N_1b_2/b_1)-2$, and the probability that the two samples share a single *b* is $P_b \approx \exp(-\Delta AIC/2-2)$. As before, we regard $\log_{10}P_b \leq -1.3$ ($P_b \leq 0.05$) as a significant difference and $\log_{10}P_b \leq -2.6$ as a highly significant one[11,16,32], and we required in addition that the absolute difference $|b_2 - b_1|$ exceed the sum of the two bootstrap uncertainties[47].

## *p*-value and postseismic-slip analyses of the 2019 study

The identification published in 2019 (ref. 29) rested on three analyses, which we summarize here because the Results refer to them. First, the change in the *b*-value was tested region by region. Nine rectangular regions were laid along the Futagawa and Hinagu FZs (Fig. 2d of previous study[29]), of which Region 6 is the red rectangle in Fig. 2, and in each region the *b*-value from 1 Jan. 2017 to 11 Mar. 2019 was compared with that from 1 Jan. 2000 to immediately before the 2016 sequence by the test of Utsu described above. The *b*-value rose in all regions except Region 6, where it fell by 0.11 ($\log_{10}P_b = -2.75$); the rises were significant except in Regions 7 and 9, at the southern end of the Hinagu FZ, which lie farthest from the 2016 ruptures. This significance was also confirmed with the bootstrap criterion described above[47]. Second, the *p*-value of the OU law, $\lambda(t)=k(c+t)^{-p}$, where $\lambda$ is the rate of earthquakes at time *t* after the *M*7.3 earthquake and *k*, *c*, and *p* are constants[35], was estimated in each region by maximum likelihood[50], with uncertainties by bootstrapping, for earthquakes above cutoff magnitudes of 1.5, 2.0, 2.5 and 3.0 from the *M*7.3 earthquake to 11 Mar. 2019; estimates from 20 or fewer earthquakes were disregarded. *p* was 1–1.4 in Regions 1–6, close to 1 in Region 6, and below 1 in Regions 7–9, and the difference between the northern and southern regions of the Hinagu FZ was significant given the uncertainties. Under the rate- and state-dependent model of seismicity[36], $p>1$ corresponds to stress that is falling rapidly, $p<1$ to stress that is falling slowly and $p\approx1$ to stress that is constant or increasing. Third, the postseismic deformation model of Pollitz et al.[25] places afterslip on planes that represent the Futagawa FZ and the northern section of the Hinagu FZ and end near the southern limit of the *M*5-class events of 2016. Afterslip dominates the postseismic relaxation near these planes and viscoelastic relaxation of the lower crust and mantle dominates farther away, so stress on the planes was falling rapidly and stress on the southern section of the Hinagu FZ, on which neither coseismic nor postseismic slip occurred, was falling slowly; at the boundary between the two, which is Region 6, the stress was constant or increasing[29].

**Fault models**

For the 2026 earthquake, we used two preliminary source models that were available at the time of writing; both may be revised. The first is the slip distribution obtained by the NIED by multi-time-window linear inversion of the S-wave portion (0.08–0.8 Hz) of three-component velocity records at 15 stations of K-NET (Kyoshin Network), KiK-net (Kiban-Kyoshin Network) and F-net (Full Range Seismograph Network of Japan), on a fault plane of strike 211° and dip 76° taken from the F-net moment tensor solution, 28 km long and 16 km wide, with the Hi-net (High Sensitivity Seismograph Network Japan) hypocentre as the point of rupture initiation[39]. The seismic moment is $2.2\times10^{19}$ Nm (moment magnitude $M_w$6.8) and the maximum slip is 5.2 m.

The second is the slip distribution of the GSI[38]. Two fault planes, F1 (strike 210°, dip 61°) and F2 (strike 225°, dip 44°), whose geometry was first estimated by a Markov chain Monte Carlo inversion, were subdivided into subfaults, and the slip on them was estimated from the displacements observed by GNSS and SAR. The GNSS displacements are the differences between positions averaged over 21–28 Jul. and over 29 Jul.–2 Aug. 2026, and therefore include those of the *M*6.1 event and, in part, of the *M*5.8 event. The moment magnitude is $M_w$6.80 and the maximum slip about 2.0 m.

Neither model was available in numerical form, so we reconstructed each as a set of rectangular fault patches from the published figures (Fig. 3 of previous study[39] and Fig. 1 of previous study[38]). For the NIED model[39], the slip on each of 112 patches (2×2 km; 14 along strike and 8 down dip) was read from the colour scale of the published slip distribution on the fault plane. The top of the plane, at a depth of 1.8 km, was read from the depth axis of the same figure, and the plane was located by placing the point of rupture initiation at the JMA hypocentre, because the Hi-net hypocentre used by NIED is not given in the report. For the GSI model[38], the slip on each of 126 patches (2.5 km along strike and 2.9 km down dip; 70 on F1 and 56 on F2) was read in the same way, and its rake from the direction of the slip vector drawn on the patch in the same figure. F2 was placed so that its northeastern upper corner coincides with the southwestern upper corner of F1, as in the published map. The reconstructed GSI model reproduces the published moment magnitude to within 0.02 when the moment is computed with the rigidity used by GSI (40 GPa); the NIED report does not give the rigidity, so the NIED model was checked with the slip vectors instead.

In both figures, the slip on each patch is also drawn as a vector, from whose direction we read the rake of the patch; where the vector was too short to read, the rake of the nearest readable patch in the same column was used. In the GSI model, the rake rotates along strike, from about 180° in the northeastern part of F1 to about −164° at its southwestern end, and from about −162° to −145° on F2 (slip-weighted means −173° and −156°). In the NIED model, it differs from patch to patch by up to about 50° from the F-net value, with reverse components on some parts of the plane and normal components on others, but its slip-weighted mean, −169°, is close to the F-net value (−170°). The lengths of the vectors, which are proportional to slip, are consistent with the slips read from the colour scales; in the NIED figure, which has a 6 m reference vector, they agree with the slips to within 2% in the median.

**Coulomb stress change**

The Coulomb stress change on a receiver fault, that is, the change in Coulomb failure stress, is $\Delta CFS = \Delta\tau + \mu' \Delta\sigma_n$, where $\Delta\tau$ is the change in shear stress in the slip direction of the receiver fault, $\Delta\sigma_n$ is the change in normal stress (positive for unclamping) and $\mu'$ is the effective coefficient of friction. A typical value, $\mu'=0.4$, was used. Positive ΔCFS brings the receiver fault closer to failure. Stress changes were computed in an elastic half-space[51] with typical values of Poisson's ratio (0.25) and Young's modulus ($8\times10^{10}$ Pa), which give a shear modulus of $3.2\times10^{10}$ Pa, using Coulomb 3.4 (ref. 52). The GSI and NIED models[38,39], reconstructed as described above, were used as the sources. The stress changes were resolved at a depth of 10 km, a typical depth of the seismicity along the Hinagu FZ (Fig. 3b), on receiver faults of two types, each with the corresponding model as the source (Fig. 5): the GSI-F2 type, with the strike and dip of F2 (strike 225°, dip 44°) and the slip-weighted mean of its rakes (−156°) in the slip distribution of GSI[38], and the NIED type, with those of the F-net moment tensor solution (strike 211°, dip 76°, rake −170°)[53], from which the plane of the NIED model was taken[39]. The stress changes were also resolved on the plane conjugate to each type, that is, the other nodal plane of the double-couple mechanism given by its strike, dip and rake: the normal of the conjugate plane is parallel to the slip vector on the first plane, and its slip vector is parallel to the normal of the first plane[54]. The conjugate planes are strike 117°, dip 74°, rake −49° for the GSI-F2 type, and strike 119°, dip 80°, rake −14° for the NIED type (Supplementary Fig. 2). Because calculated stress changes on and very near the faults that slipped are not reliable[29,43], only the values beyond the southwestern end of the slipped area are interpreted.

**Data availability**

The JMA earthquake catalogue is available at https://www.data.jma.go.jp/eqev/data/daily_map/index.html (accessed 5 Oct. 2026). The F-net moment tensor catalogue is available at https://www.fnet.bosai.go.jp/ (accessed 5 Oct. 2026). The preliminary source models and evaluations of the 2026 earthquake are contained in the materials released by ERC[28] dated 29 Jul. and 12 Aug. 2026 (https://www.static.jishin.go.jp/resource/monthly/2026/20260728_kumamoto_1.pdf and https://www.static.jishin.go.jp/resource/monthly/2026/20260728_kumamoto_2.pdf, accessed 5 Oct. 2026). The long-term evaluation of the Futagawa and Hinagu FZs[19], including the location map from which the faults in the Yatsushiro Sea were digitized, is available at https://www.jishin.go.jp/regional_seismicity/rs_katsudanso/f093_futagawa_hinagu/ (accessed 5 Oct. 2026). Active fault data provided in the software Coulomb were used[55,56], to which we added the faults in the Yatsushiro Sea digitized from that map[19] with a positional accuracy of about 0.3 km. The *b*-value maps and depth sections, the reconstructed source models, the digitized fault traces and the Coulomb stress calculations generated in this study are deposited at Figshare[57] (https://doi.org/10.6084/m9.figshare.34078323), where the data behind each figure are provided together with the code that draws it. The deposit will be made public, and the DOI activated, upon acceptance; during peer review it can be accessed through the private link https://figshare.com/s/9c4fd57e3417b62cf0f5.

**Code availability**

The EMR method[44] was implemented in ZMAP[58], available at https://github.com/CelsoReyes/zmap7. Coulomb stress changes used for Fig. 5 and Supplementary Fig. 2 were computed with Coulomb 3.4 (the graphic-rich deformation and stress-change software)[52,55,56], available for download at https://temblor.net/coulomb/. The depth sections of Figs. 3b–e and 4 and of Supplementary Fig. 1 were drawn with MATLAB R2024b (The MathWorks), and the maps of Figs. 1, 2 and 3a with the Generic Mapping Tools version 6.4.0[59]. The scripts that produce the figures are deposited, together with the data they use, at Figshare[57] (https://doi.org/10.6084/m9.figshare.34078323; see Data availability for access during peer review).

**Acknowledgements**

We thank JMA for the earthquake catalogue and T. Nagao for useful discussions. Research by K.Z.N., J.I., and M.K. was partially supported by Japan's Ministry of Education, Culture, Sports, Science and Technology (MEXT) under the Second Earthquake and Volcano Hazards Observation and Research Program (Earthquake and Volcano Hazard Reduction Research); and the SECOM Science and Technology Foundation (FY2023 Special Research Grant). Research by J.B.R. was supported in part by the Southern California Earthquake Center (SCEC) under grant no. SCON-00007927 to UC Davis, and by the John LaBrecque Fund, a generous gift to the University of California, Davis.

**Author contributions**

K.Z.N. and J.B.R. conceptualized the research. K.Z.N., D.I., and J.I. contributed to the development and refinement of the methodology. K.Z.N. and M.K. contributed to the data curation. K.Z.N. analysed the data and prepared the figures. K.Z.N. and J.B.R. supervised the research. K.Z.N. wrote the original draft of the paper. All authors substantively reviewed and edited the paper and approved the final version.

**Competing interests**

The authors declare no competing interests.

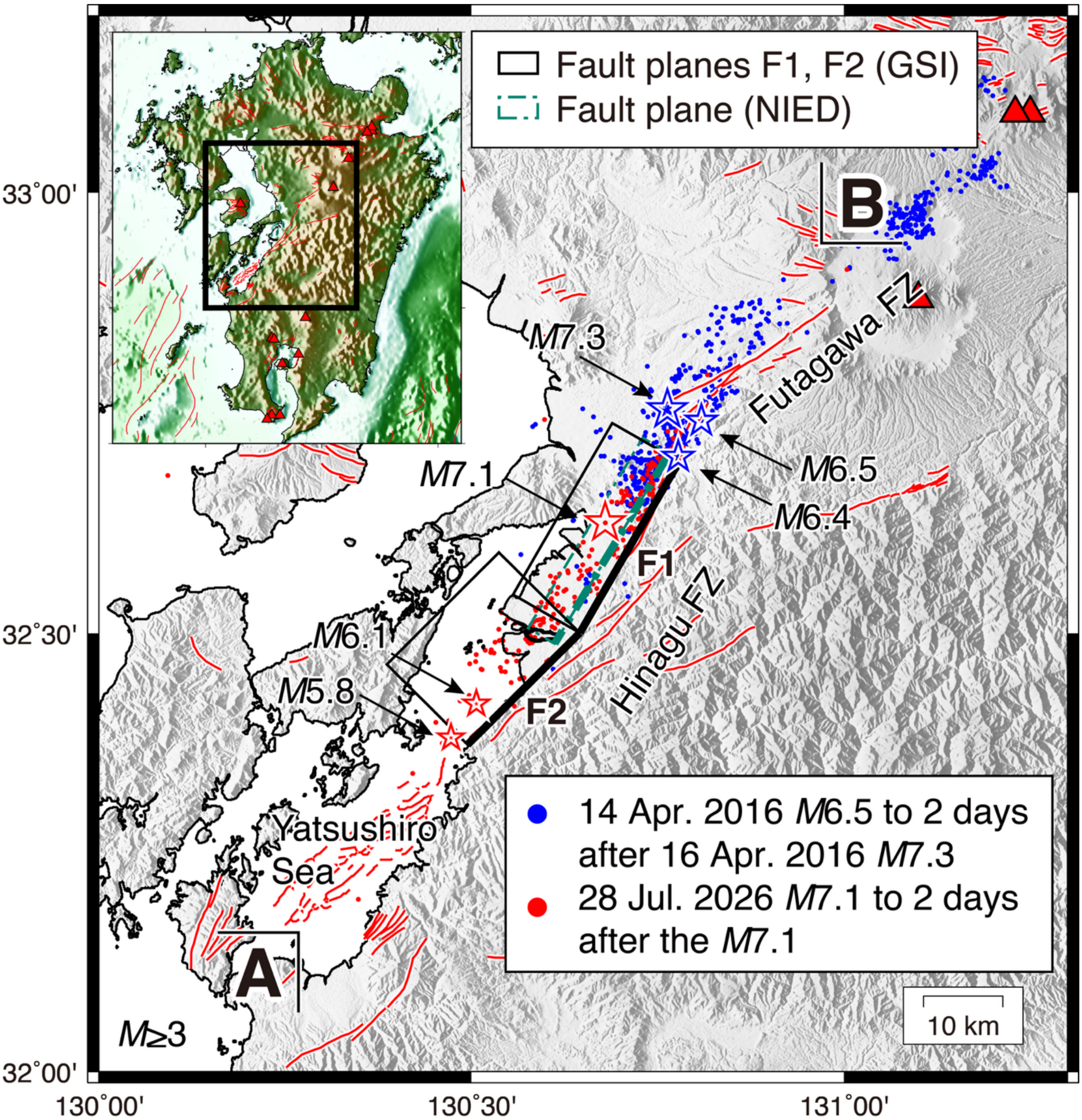

Figure 1. Map view of the Kumamoto region. Blue stars show three large earthquakes of the 2016 Kumamoto sequence: *M*6.5 (14 Apr. 2016), *M*6.4 (15 Apr. 2016), *M*7.3 (16 Apr. 2016). Blue dots show $M\geq3$ events (depths 25 km or shallower) during the period from the *M*6.5 earthquake to 2 days after the *M*7.3 earthquake. Red stars are the same as the blue stars for three large earthquakes of the 2026 Kumamoto sequence: *M*7.1 (28 Jul. 2026), *M*6.1 (28 Jul. 2026), and *M*5.8 (29 Jul. 2026). Red dots are the same as blue dots for the period of 2 days after the *M*7.1 earthquake. Red triangles show active volcanoes. Red lines mark active faults. Black and green rectangles are the surface projections of the fault planes of the slip

distributions of GSI[38] and NIED[39] used in Figs. 3d, e and 5; thick lines mark their upper edges. Cross-sectional views in Figs. 3 and 4 and Supplementary Fig. 1 extend from A to B. The zoomed inset is a map showing the study region (black rectangle) in the Kyushu district.

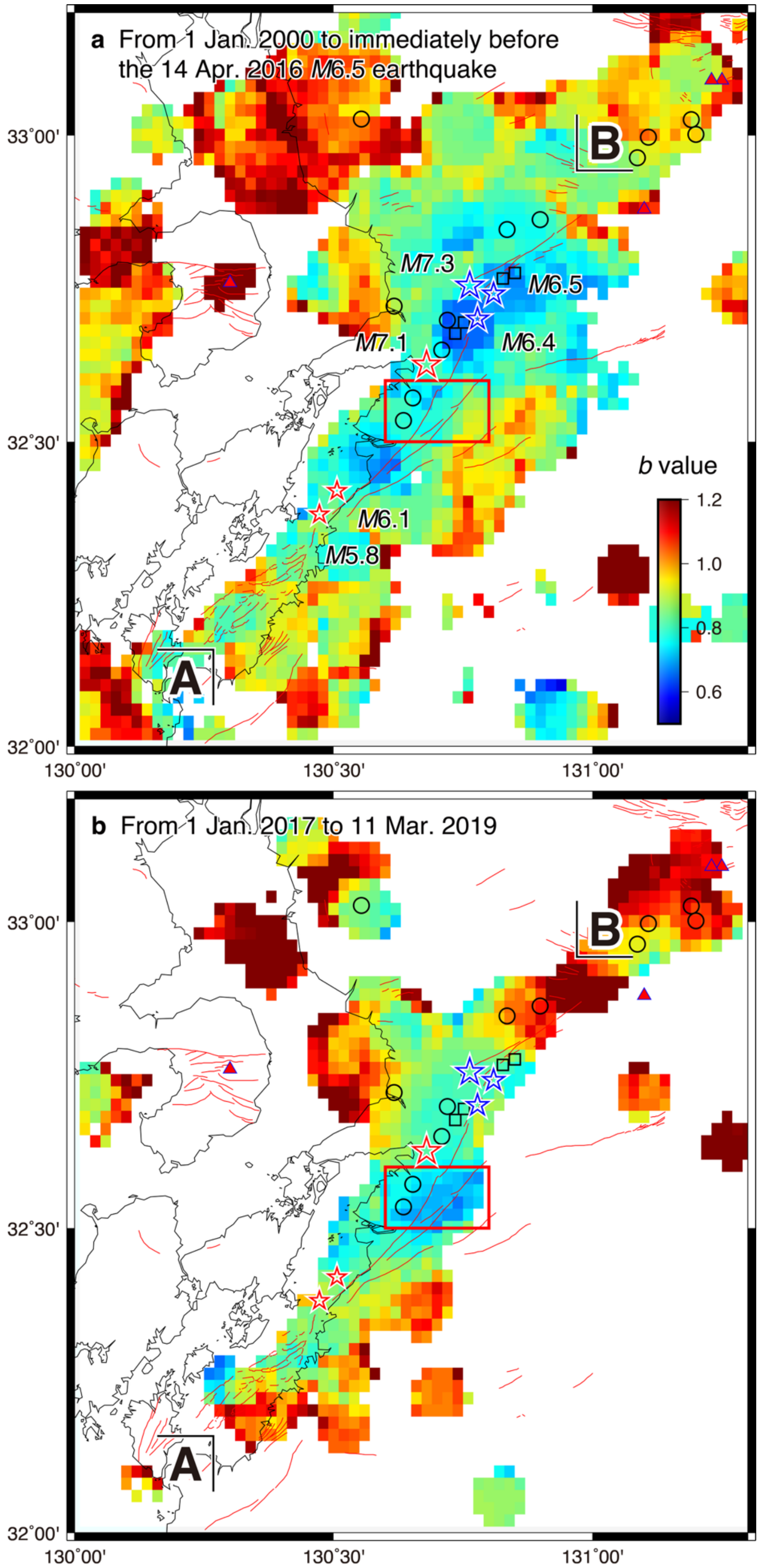

Figure 2. Map views of $b$-values in the Kumamoto region. The $b$-values in **a** and **b** were reproduced from previous studies[29,30]. **a**, $b$-values based on seismicity during the period from 1 Jan. 2000 to immediately before the 2016 Kumamoto sequence. Red rectangle shows the region of 130.6–130.8°E and 32.5–32.6°N, indicated by previous study[29]. For the explanation of blue stars, red stars, red lines, red triangles, A, and B, see the caption of Fig. 1. Squares show $M\geq5$ events during the period between the 2016 *M*6.5 and *M*7.3 earthquakes. Circles are the same as squares for the period between the 2016 *M*7.3 earthquake and the 2026 *M*7.1 earthquake. **b**, Same as **a** for *b*-values based on seismicity during the period from 1 Jan. 2017 to 11 Mar. 2019.

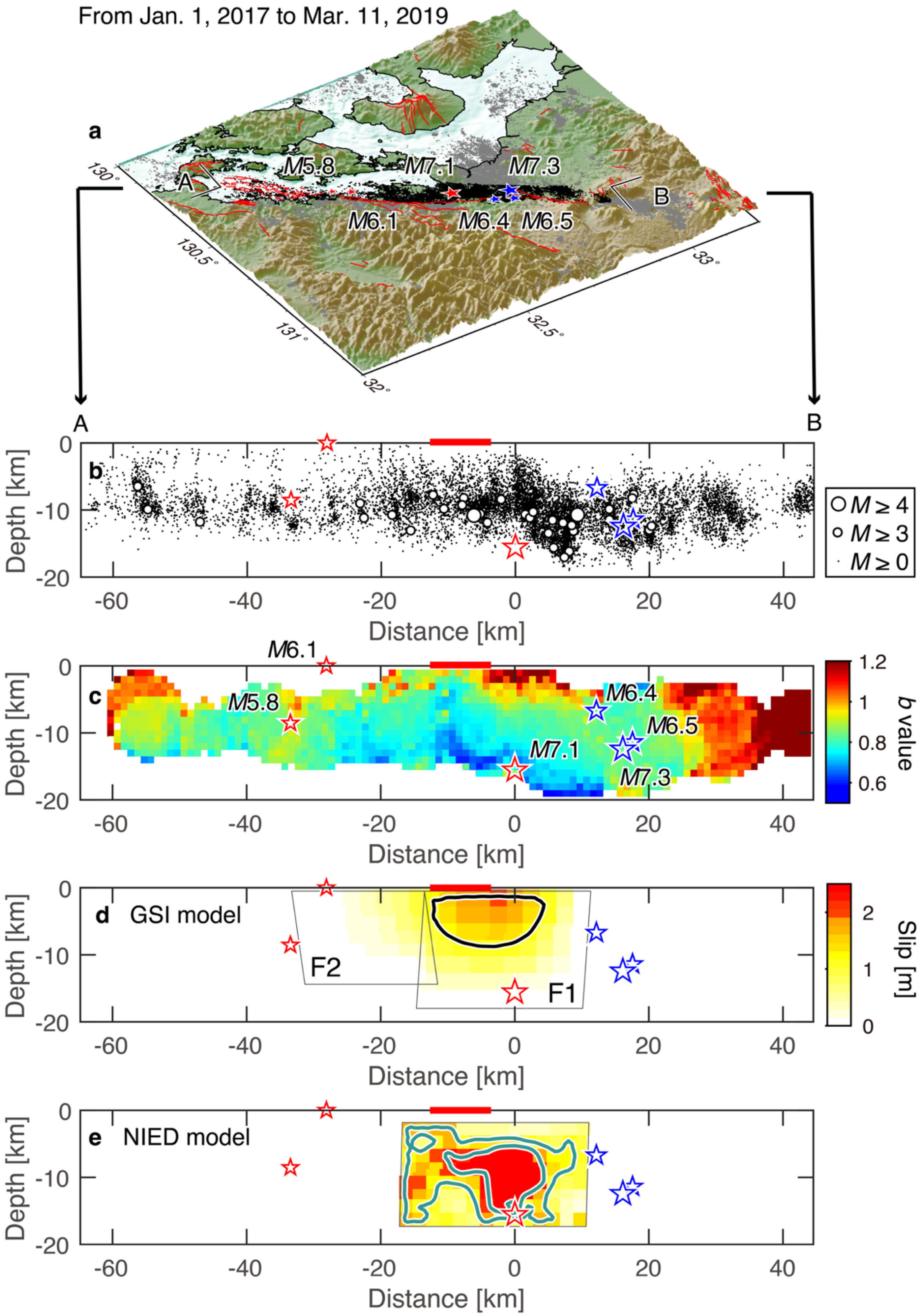

Figure 3. Cross-sectional views of the Hinagu FZ. **a**, Topographic map of the region around the Hinagu FZ. Seismicity ($M \geq 0$) during the period from 1 Jan. 2017 to 11 Mar. 2019 in this region is shown by grey dots. This period is the same as that used in Fig. 2**b** and as that used in previous study[29]. Black dots indicate seismicity within a width of 10 km from A to B. This seismicity was used to create the cross-sectional views in **b** and **c**. For the explanation of stars and red lines, see the caption of Fig. 1. **b, c,** Cross-sectional views of seismicity (1 Jan. 2017–11 Mar. 2019) and *b*-values. Red bar corresponds to the red rectangle in Fig. 2. **d,** Preliminary slip distribution of the 2026 Kumamoto sequence obtained by GSI[38]. Contour line shows slip of 1.2 m. Parallelograms indicated by F1 and F2 show the fault planes (black rectangles) in Fig. 1. **e**, Same as **d** for that obtained by NIED[39]. Contour lines show 1.2 and 2.4 m.

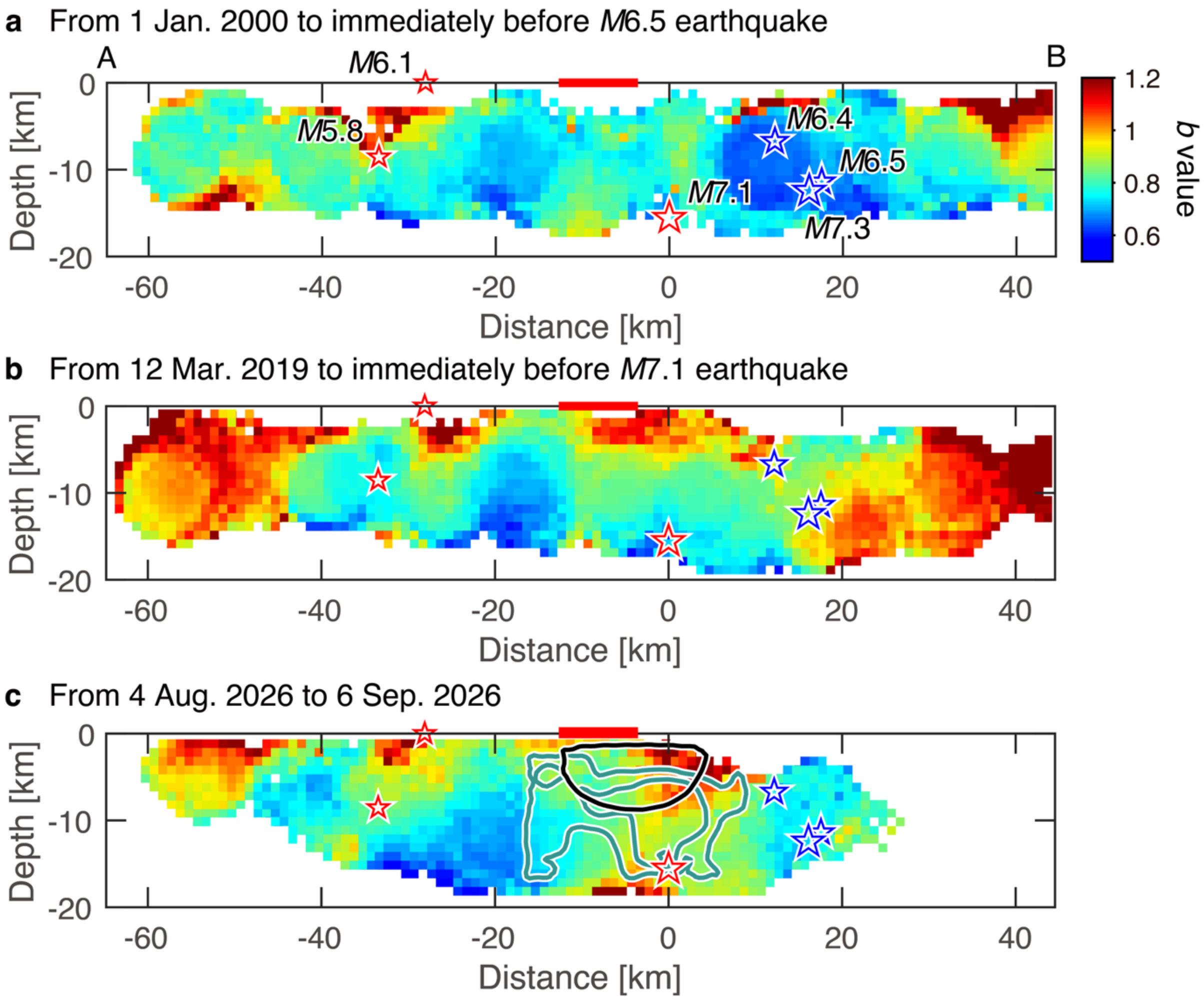

Figure 4. *b*-values mapped along the cross-section. **a**, Same as Fig. 3**c** for the period from 1 Jan. 2000 to immediately before the 2016 *M*6.5 earthquake. This period is the same as that used in Fig. 2a and that used in previous studies[29,30]. For the explanation of red bar and stars, see the caption of Fig. 3. **b**, Same as **a** for the period from 12 Mar. 2019 to immediately before the *M*7.1 earthquake. **c**, Same as **a** for the period after the 2026 *M*7.1 earthquake (4 Aug.–6 Sep. 2026). For the explanation of slip contours, see the caption of Fig. 3.

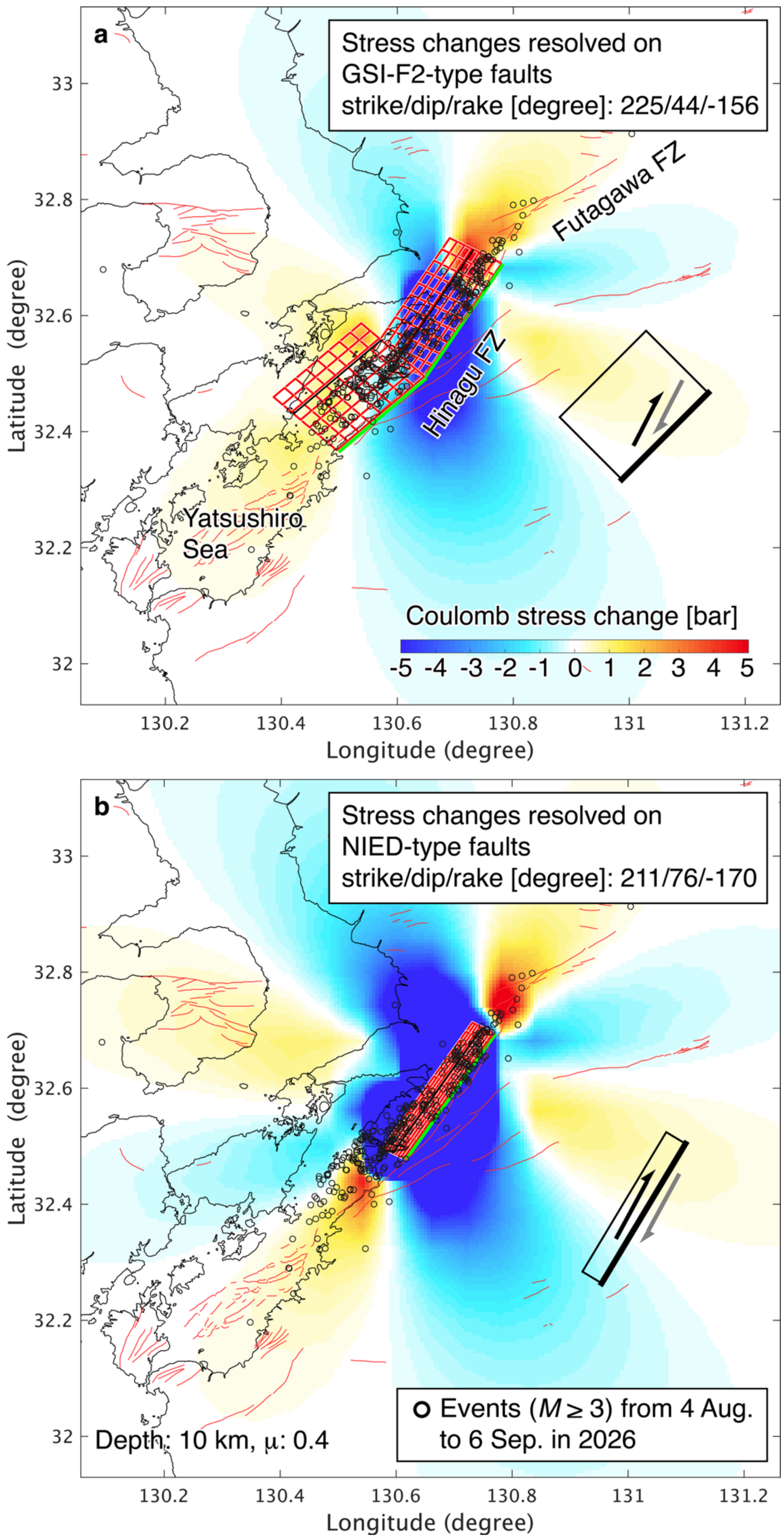

Figure 5. Coulomb stress changes imparted by the 2026 *M*7.1 earthquake. The changes were calculated at a depth of 10 km with an effective coefficient of friction of 0.4. **a**, Stress

changes resolved on faults of the GSI-F2 type (strike 225°, dip 44°, rake −156°), with the GSI model[38] as the source. The red grid shows the fault patches of the source model. **b,** Same as **a** for faults of the NIED type (strike 211°, dip 76°, rake −170°), with the NIED model[39] as the source. Grey circles show earthquakes of $M\geq3$ from 4 Aug. to 6 Sep. 2026. For the explanation of red lines, see the caption of Fig. 1. The symbol in each panel shows a receiver fault in map view: the rectangle is the surface projection of the fault plane, with its upper edge drawn as a thick line, and the black and grey arrows show the slip of the hanging wall and the footwall, respectively, projected onto the horizontal plane. On each source fault, the green line is its surface trace, the plane extended up-dip to the surface, and the black line is its intersection with the depth of 10 km.

# Supplementary Information

# Advance identification of where the 2026 Kumamoto earthquake would begin

**K. Z. Nanjo, J. B. Rundle, D. Iwata, J. Izutsu, M. Kamogawa**

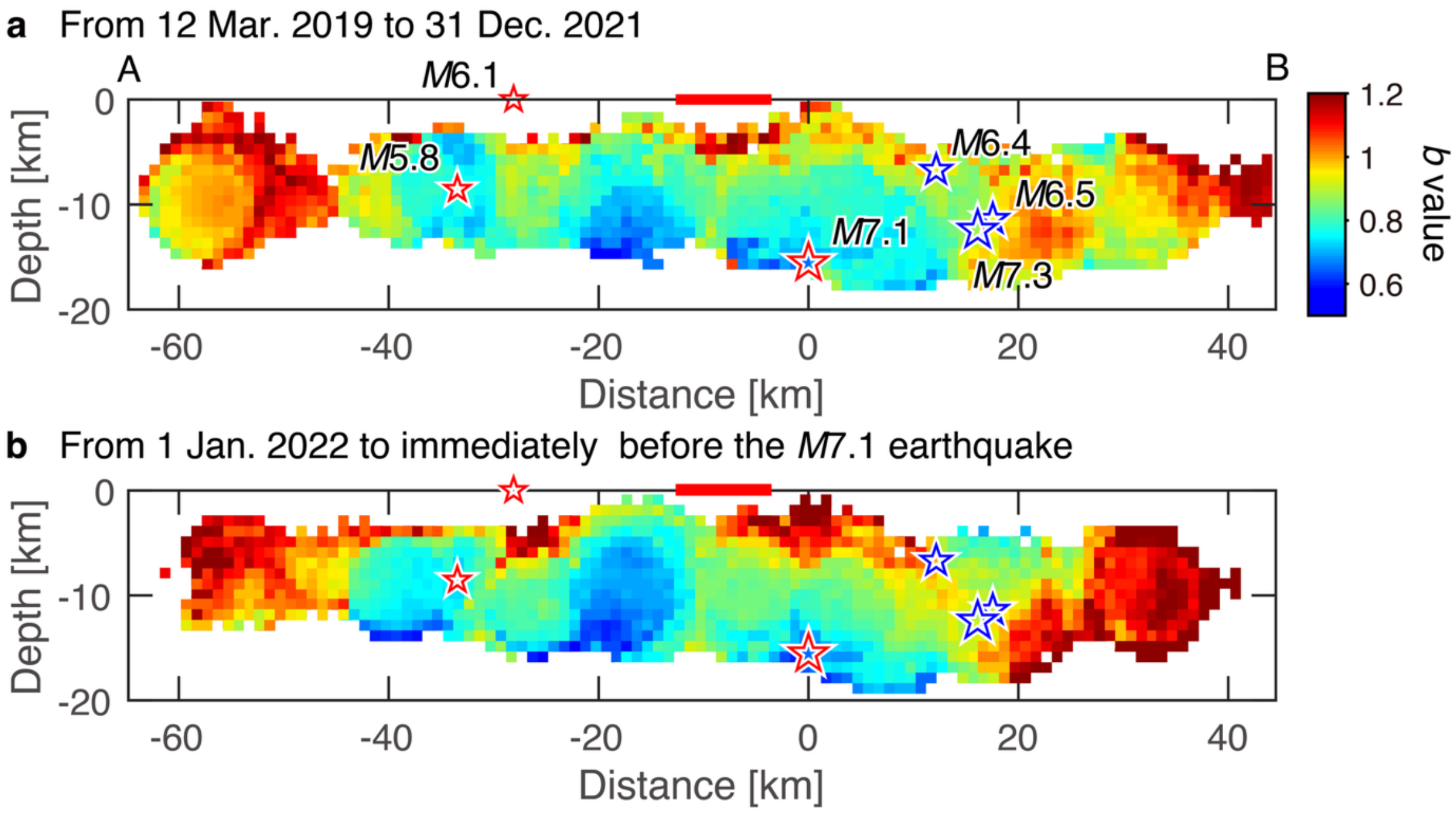

Supplementary Figure 1. Same as Fig. 4**b**, but we divided the period in Fig. 4**b** into two sub-periods: **a**, from 12 Mar. 2019 to 31 Dec. 2021 and **b**, from 1 Jan. 2022 to immediately before the 2026 *M*7.1 earthquake.

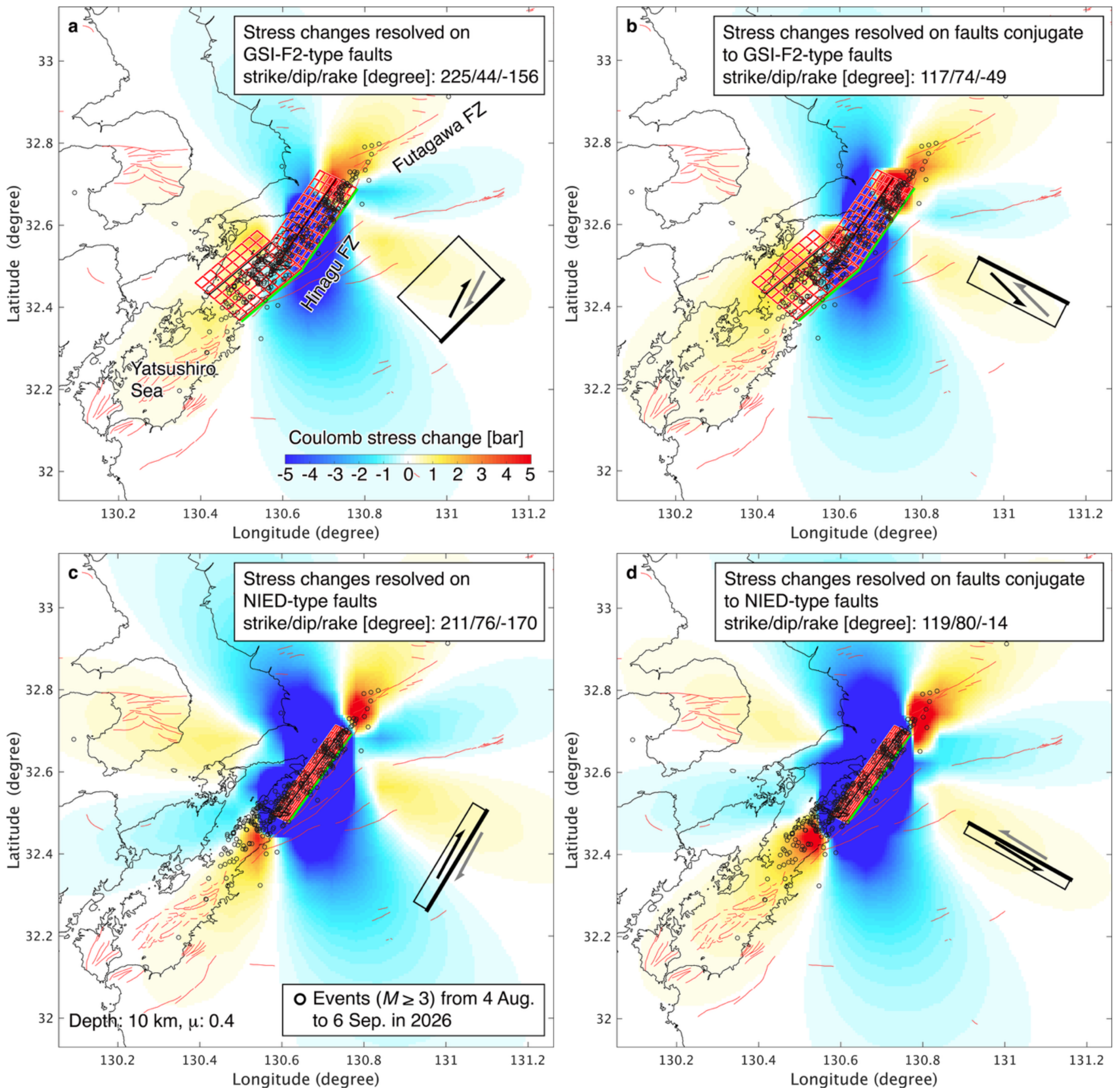

Supplementary Figure 2. Same as Fig. 5 for conjugate faults. **a**, Same as Fig. 5**a**. **b**, Same as **a** for faults conjugate to those in **a**. **c**, Same as Fig. 5**b**. **d**, Same as **c** for faults conjugate to those in **c**.